# REVEALING SUB-OPTIMALITY CONDITIONS OF STRATEGIC DECISIONS

H. KEMAL İLTER

Department of Management, Baskent University, Eskisehir Yolu 20.km, 06530, Ankara, Turkey

*Draft version June 30, 2011*

## ABSTRACT

Conceptual view of fitness and fitness measurement of strategic decisions on information systems, technological systems and innovation are becoming more important in recent years. This paper determines some dynamics of fitness landscape which are lead to termination of decision makers' research before reaching the global maximum in strategic decisions. These dynamics are specified according to management decision making models and supported with simulation results. This article determines simulation results by means of "Fitness Value" and "Probability of Optimality". Correlation between these two concepts may be remarkable according to revealing optimal values in innovative and research-based decision making approaches beside sub-optimal results of traditional decision making approaches.

## 1. INTRODUCTION

Fitness landscape theory is becoming to use for answering the search of developing species which desire to reach highest peak on the potential gene space in the field of evolutionary biology (Wright, 1932; Gillespie, 1984). Development of the cost landscapes in related solution space is used as an approach to combinatorial optimization problems' solutions (Holland, 1975; Kirkpatrick et al., 1983; Palmer, 1988) in computer engineering and operations research fields.

In recent years, different approaches are used in various fields of social sciences like organizational change (Beinhocker, 1999; McKelvey, 1999; Reuf, 1997), evolution of social structures (Levinthal, 1996), innovation networks (Frenken, 2000 and 2006), selection of appropriate technology (McCarthy and Tan, 2000; McCarthy, 2003), economic structures (Kauffman, 1993) and political systems (Kollman et al., 1992).

Introducing perspective with NK model which is simplify using fitness landscape theory in various fields is devoted to be possible in global optimum searching on a stochastic but easily controllable fitness landscape that composed of possible fitness values.

Local optimum points besides global optimums are also important in fitness landscapes (see for detailed information: Ilter, 2007; Ilter, 2008). Local optimum points can be seen peak points which isn't allow for changing possible fitness values even if alternatives of selection are changed. After all, firms can terminate their search for global or local optimum value(s) because of sub-optimal value(s) are accepted as best value(s) for them. "Why does the firm generally terminate their search on the fitness landscape before finding the local optimum value yet?" is still an unanswered question in social sciences.

Decisions on information systems, technology and innovation have a different place in firms' decision mechanism in terms of some aspects. There are various factors seems to be important which affect the selection of technological structures (production technology, information technology, etc.) in firms and the decision behaviors of decision makers in making decision related to these technologies. These behaviors can be revealed by affecting the optimality of firm's final decision and the inner-fitness of decision maker in a large extent. It is possible to say that organization properties and factors in organization hierarchy effect decisions about information systems, technology and innovation management which can be concluded as a part of complex systems. Decisions except optimal ones (sub-optimal decisions) couldn't recognized while design of decision mechanism in organization targets the optimal decision in some conditions.

## 2. METHOD

In this article, we try to determine some important dynamics that may cause of termination of firms' searching action of optimality even if they couldn't reach the local optimum value. These dynamics are considered together with some factors of organizational decision-making models and then supported with simulation results to reveal sub-optimality conditions.

Various scenarios have reviewed by using NK fitness landscape theory for determining sub-optimality conditions that could be in decisions which are related to information systems, technology and innovation. Scenarios are developed as some conditions which are include one decision maker, various numbers of subordinates and various numbers of decisions (Table 1). The state of optimality divergence and the state of sub-optimality toleration of decision maker emphasized in this article are supported by simulations' results after statistically acceptable number of runs. As an example, scenario L07 is reflecting a case which includes two subordinates (subordinate A has to make a decision and subordinate B has to make three decisions) and a passive decision maker (has to make a decision of subordinates' decision combination).

Several runs of various scenarios are inspected with using NK fitness theory for determining sub-optimality states of decisions which are related to information systems, technology and innovation.

**Table 1: Simulation Scenarios**

| Scenario Code | L01/L02 | L03/L04 | L05/L06 | L07/L08 | L09/L10 | L11/L12 | L13/L14 |
|---|---|---|---|---|---|---|---|
| Decision Maker's Mode | Passive and Active | | | | | | |
| Number of Subordinate | 2 | | | | | | |
| Number of decision that assigned to each subordinate | 1-1 | 1-2 | 2-2 | 1-3 | 3-3 | 2-4 | 1-5 |
| Relation between decision components | K = N - 1 (high complexity)* | | | | | | |
| Number of decision alternative that is forwarded from subordinate to decision maker | 1 | | | | | | |
| Number of decision level (hierarchical structure) | 1 | | | | | | |
| Level of the relation between decision components | affected equally** | | | | | | |

\* K perimeter in NK model is a value that specifies effects of decision components on each other. In this article, we assume that every decision component is affecting every other decision component in some way and there is the highest possible complexity level while decision alternatives are evolving in simulation scenarios.

\*\* We assume that decision components are affecting each other in equal manner. On the other hand, there is an analysis about scenarios of L03 and L04 for showing non-equal effects divergence.

## 3. RESULTS

In simulation, global optimum points that on the fitness landscape (space of decision alternatives) are determined by generation of a landscape which includes fitness values of each scenario. Probability of achieving global optimum in final decision of the decision maker after evaluation of decision alternatives can be defined as the term of "Probability of Optimality". Sub-optimality is appeared if decision's probability of optimality descents below 100%. Relationship between fitness value of the final decision of the decision maker and global optimum value on the fitness landscape can be defined as "Fitness Rate" and it can be recognized as of success factor of decision maker's final decision. Successes of decisions can be determined by inspection on global optimum-decision's fitness value bias for the decision that can't achieve optimality but sub-optimality. (Figure 1)

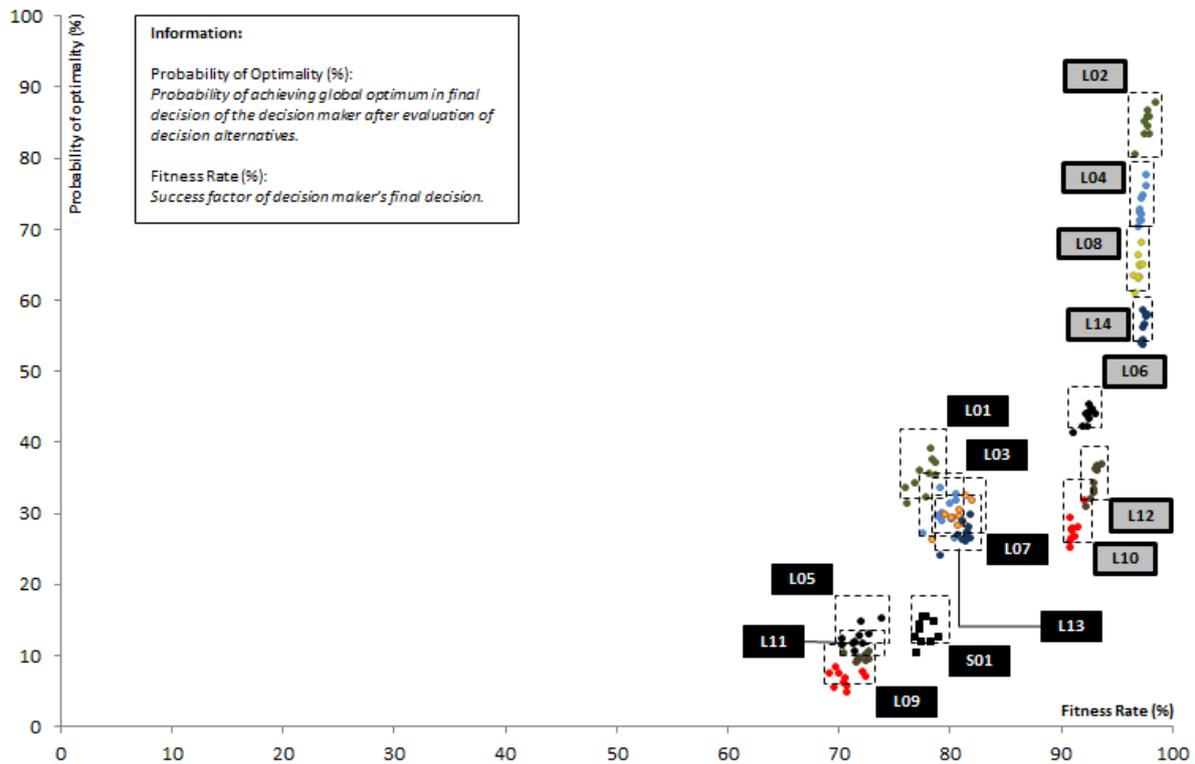

Figure 1: Relationship Between Probability of Optimality and Fitness Rate in Various Scenarios

4. CONCLUSION

Probabilities of optimality and fitness values of each scenario are correlated each other during the analysis of simulation results. These correlations show that decisions of active decision makers (scenario code in light color) are more efficient than decisions of passive decision makers (scenario code in dark color) in terms of the fitness value. On the other hand, there are no optimal values for each of all decisions and noticed that these decisions are not optimal but sub-optimal. Addition to this determination, decisions of active decision makers have higher values than decisions of passive decision makers in terms of the probability of optimality. Success of decision maker is limited from the point of view of the probability of optimality in despite of these two settings are stated that the active decision makers are more successful than passive decision makers.